\documentclass[preprint,showpacs,preprintnumbers,amsmath,amssymb]{revtex4}

\usepackage{graphicx}
\usepackage{dcolumn}
\usepackage{bm}
\usepackage{gensymb}
\usepackage{mathtools}
\usepackage{float}
\usepackage{chngcntr}
\usepackage{soul}
\usepackage[title]{appendix}
\usepackage{xcolor}
\usepackage[per-mode=symbol]{siunitx}
\graphicspath{ {./Images/} }
\usepackage{amsmath}
\usepackage{mathrsfs} 
\usepackage{siunitx} 
\newcommand{\dd}[1]{\mathrm{d}#1}
\DeclareMathOperator\erf{erf}

\begin{document}

\preprint{}

\title{Controlling capillary fingering \\using pore size gradients in disordered media}

\author{Nancy B. Lu}
  \altaffiliation{Chemical and Biological Engineering, Princeton University, Princeton, NJ 08544}
\author{Christopher A. Browne}
  \altaffiliation{Chemical and Biological Engineering, Princeton University, Princeton, NJ 08544}
 \author{Daniel B. Amchin}
  \altaffiliation{Chemical and Biological Engineering, Princeton University, Princeton, NJ 08544}
\author{Janine K. Nunes}
  \altaffiliation{Mechanical and Aerospace Engineering, Princeton University, Princeton, NJ 08544}
\author{Sujit S. Datta}
  \altaffiliation{Chemical and Biological Engineering, Princeton University, Princeton, NJ 08544}
   
\date{\today}

\begin{abstract}
Capillary fingering is a displacement process that can occur when a non-wetting fluid displaces a wetting fluid from a homogeneous disordered porous medium. Here, we investigate how this process is influenced by a pore size gradient. Using microfluidic experiments and computational pore-network models, we show that the non-wetting fluid displacement behavior depends sensitively on the direction and the magnitude of the gradient. The fluid displacement depends on the competition between a pore size gradient and pore-scale disorder; indeed, a sufficiently large gradient can completely suppress capillary fingering. By analyzing capillary forces at the pore scale, we identify a non-dimensional parameter that describes the physics underlying these diverse flow behaviors. Our results thus expand the understanding of flow in complex porous media, and suggest a new way to control flow behavior via the introduction of pore size gradients.

\end{abstract}

\pacs{Valid PACS appear here}
\maketitle
\bibliographystyle{ieeetr}
\section{\label{sec:level1}Introduction}
\noindent Drainage is the process by which a non-wetting fluid displaces a wetting fluid from a porous medium. This phenomenon is ubiquitous: it arises in diverse settings including groundwater contamination, oil migration, gas venting from sediments, CO$_2$ sequestration, mercury porosimetry, soil drying, liquid infusion into porous membranes, and oxygen accumulation within polymer electrolyte membranes \cite{cueto2008nonlocal, bazyar2018liquid, sahimi2011flow, berg2012stability, bethke1991long, dandekar2013petroleum, benson2008carbon, macminn2010co, saadatpoor2010new, bandara2011pore, neufeld2009modelling, bear2010modeling, dawson1997influence, levy2003modelling, kueper1991two, lee2016porous, carmo2013comprehensive}. The ability to accurately predict the displacement pathway of the non-wetting fluid is critically important in all of these cases \cite{sahimi2011flow, rabbani2018suppressing}. For instance, whether the fluid flows through a compact, stabilized front or a narrow and fingered channel impacts how much contaminant penetrates into an aquifer, how much oil can be recovered from a reservoir, and how much water remains in a dried soil \cite{cottin2011influence,singh2003dynamic, datta2014mobilization, datta2013drainage}.

Different displacement behaviors can arise during drainage. For a homogeneous disordered medium of uniform wettability and uncorrelated pore sizes, these behaviors are predictable using two non-dimensional parameters: the viscosity ratio $\text{M}\equiv\mu_\text{nw}/\mu_\text{w}$ and the capillary number $\text{Ca}\equiv\mu_\text{nw} (Q/A)/\gamma$, where $\mu_\text{nw}$ and $\mu_\text{w}$ are the non-wetting and wetting fluid viscosities, $Q$ is the fluid flow rate through a cross-sectional area $A$ of the medium, and $\gamma$ is the interfacial tension between the two fluids \cite{lenormand1988numerical, yortsos1997phase, xu1998invasion}. Many processes are characterized by $\text{M}>1$ and $\text{Ca}\ll1$; under these conditions, capillary forces dominate, and the non-wetting fluid cannot enter a pore of diameter $a$ until the fluid pressure reaches a threshold $\sim\gamma/a$. Therefore, the non-wetting fluid displacement proceeds one pore invasion at a time. At each time, the fluid invades the largest pore accessible to it, which is characterized by the lowest capillary pressure threshold. The flow behavior is thus determined by pore-to-pore variations in the pore size, resulting in a displacement process known as capillary fingering (CF) that is characterized by a ramified and disordered pathway \cite{lenormand1989capillary, lenormand1983mechanisms, lenormand1985invasion, mayer1965mercury, xu2008dynamics, krummel2013visualizing, toledo1994pore, mason1986meniscus, joekar2012analysis, maaloy1992dynamics, martys1991critical, xu1998invasion}. 


Many naturally-occurring and synthetic porous media are not homogeneous, however. For example, shales, sandstones, and soils are typically heterogeneous, with smooth gradients or sharp discontinuities in pore size both along and orthogonal to the fluid flow direction \cite{ringrose1993immiscible, schaetzl2015soils, ashraf2019capillary}. Sharp pore size stratification has been shown to alter the fluid pathway during drainage \cite{datta2013drainage, yokoyama1981effects, lake1981taylor, chatzis1995investigation}, yet the influence of a smooth gradient in pore sizes is still unclear. Theoretical calculations, numerical simulations, and indirect experimental evidence suggest that an applied pressure gradient can modify the fluid pathway \cite{meakin1991,meakin1992,chaouche1994,xu1998}, and a pore size gradient has been conjectured to play a similar role \cite{yortsos2001}, but this conjecture has not been directly tested in experiments. Recent investigations of viscous fingering, a distinct fluid displacement behavior that arises for $\text{M}<1$, demonstrate that gradients and spatial correlations in pore size can indeed strongly impact the geometry of the fluid pathway \cite{rabbani2018suppressing, al2012control, jackson2017stability,pihler2012suppression,biswas2018drying}. However, how a pore size gradient impacts capillary fingering remains unknown. As a result, accurate prediction of fluid displacement pathways is still elusive for many real-world applications. 

Here, we use microfluidic porous media and computational pore-network models to investigate how capillary fingering is influenced by a pore size gradient. We find that the non-wetting fluid displacement behavior depends sensitively on the direction and the magnitude of the gradient, and for a sufficiently large gradient, capillary fingering is completely suppressed. Instead, if the non-wetting fluid flows down the gradient, it propagates via a uniform, stabilized front, while if the fluid flows up the gradient, it propagates through a single, unstable, fingered channel. This behavior also depends on the relative amount of disorder in the geometry of the medium; we demonstrate that the fluid displacement can be described by a single non-dimensional parameter that quantifies the competition between a pore size gradient and pore-scale disorder. Moreover, by analyzing capillary forces at the pore scale, we develop a geometric criterion that predicts when capillary fingering is completely suppressed. Our results thereby help elucidate how diverse flow pathways can arise due to pore size gradients in disordered media.

\section{Experimental materials and methods}

\begin{figure}[htp]
\includegraphics[width=\textwidth]{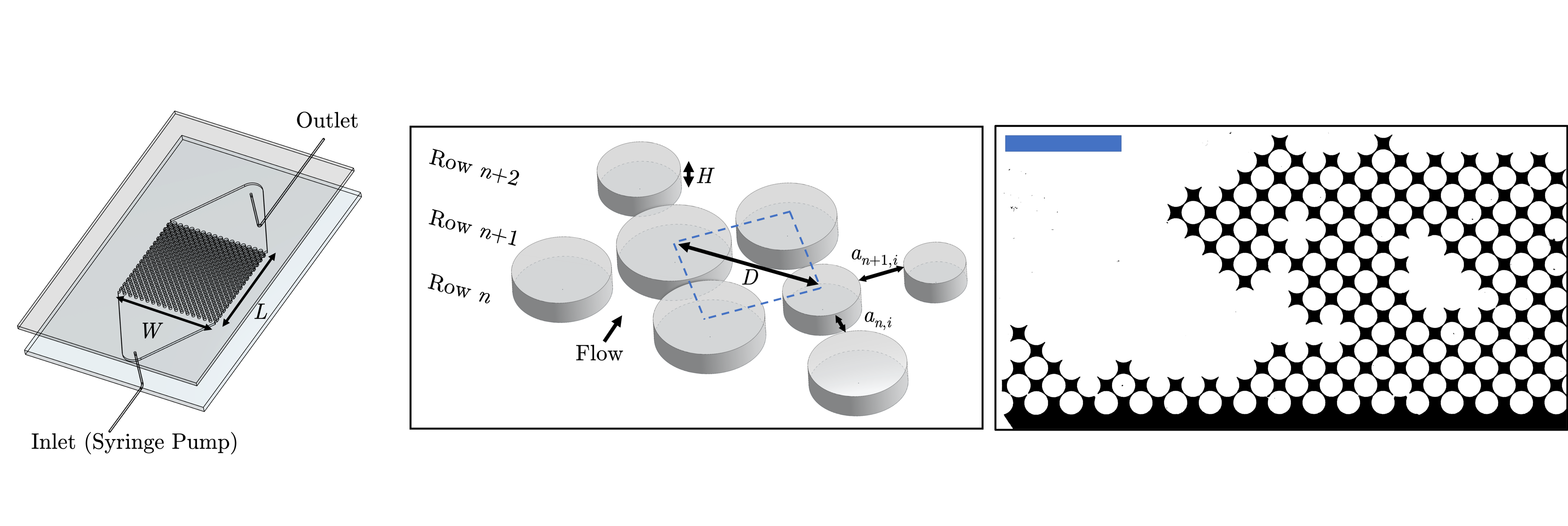}
\caption{Design of microfluidic devices with controlled pore geometries. (Left) 3D schematic of a device. The fluid flows from the inlet to the outlet. The width and length of the porous medium are labeled by $W$ and $L$, respectively. (Middle) The pillars are arranged on a square diagonal lattice; their diameters thus determine the pore diameters. The pore throat diameters are given by $a$ and are indexed by the row number and the position along the row. Schematic shows an example of a disordered porous medium with pores of variable, uncorrelated sizes. (Right) Experimental image of capillary fingering. Black shows invading non-wetting fluid, while white circles show pillars defining the solid matrix and additional white is the pore space. The scale bar represents 5 mm. Flow is from bottom to top.}
\label{fig:setup}
\end{figure}

\noindent To systematically study the influence of pore size gradients and disorder on drainage, we design polydimethylsiloxane (PDMS) microfluidic devices with controlled pore geometries. The devices are comprised of a rectangular channel of width $W=\SI{28}{\milli\meter}$ and length $L=\SI{25}{\milli\meter}$ containing a two-dimensional (2D) array of cylindrical pillars that span the channel height $H=166~\pm~$\SI{8}{\micro\meter}, where the uncertainty reflects experimental variations in device fabrication. Importantly, because $H\ll W$, the fluid flow is effectively 2D. We arrange the pillars on a square diagonal lattice with a diagonal length $D$ of \SI{1.6}{\milli\meter}, as shown in Figure~\ref{fig:setup}. The lattice has 31 rows with either 17 or 18 pillars per row. The pillar diameters determine the pore diameters: the diagonal distances between neighboring pillars represent the pore throats. We denote the diameters of the pores throats in row $n$ by $a_{n,i}$, where $i$ indexes the pores along the row, and the mean diameter by $\bar{a}_{n}$. Our experiments explore the range $\SI{111}{\micro\meter}\leq \bar{a}_{n}\leq\SI{621}{\micro\meter}$, with an uncertainty of $\pm$ \SI{1}{\micro\meter} due to experimental variability in device fabrication. We also introduce a triangular area at the inlet of each device to ensure that the flow is uniform before it reaches the first set of pillars \cite{rabbani2018suppressing}. 

We fabricate the devices using standard soft lithography. First, we design silicon molds for the devices using photolithography with a positive photoresist and deep reactive ion etching. We then cast the devices using PDMS (Dow Corning Sylgard 184), with a cross-linker to elastomer ratio of 1.5 to 10. We heat the castings for 20 minutes at \SI{150}{\celsius} and treat them for 30 seconds using a Corona treatment. Finally, we attach a glass slide to each casting and leave them to bond further overnight at  \SI{65}{\celsius}. The finished PDMS surfaces are hydrophobic and oleophilic, as previously described \cite{rabbani2018suppressing}; our images (e.g. in Figs. 2-3) indicate that interactions with the PDMS primarily dictate the mean curvature of the fluid-fluid interface in the experiments. 

Prior to each experiment, we saturate the device with the wetting fluid, silicone oil of viscosity $\mu_\text{w}$ = \SI{5}{\milli\pascal\second}. We then inject the non-wetting fluid, 76.5 vol$\%$ glycerol in water colored with food dye; the interfacial tension between the wetting and non-wetting fluids is $\gamma\approx~$\SI{30}{\milli\newton\per\meter}. The non-wetting fluid has viscosity $\mu_\text{nw}$ = \SI{50}{\milli\pascal\second}; therefore, our experiments probe $\text{M}=10$. We use a Harvard Apparatus Pump 11 Elite syringe pump to impose a constant injection flow rate $Q=0.003$ mL/h, which corresponds to $\text{Ca}\equiv\mu_\text{nw} (Q/WH)/\gamma\approx 3\times10^{-7}$, well in the capillary fingering regime for homogeneous disordered media. Our experiments confirm this expectation: we observe that the non-wetting fluid invades the pore space sequentially, one pore invasion at a time. 

As the non-wetting fluid flows, we image the evolution of the flow pathway using a mounted digital camera taking 4K resolution images between 1 and 2 frames per minute. To characterize the structure of the resulting displacement pathway, we use the binarized images to determine the fraction of the pore space volume that is occupied by the non-wetting fluid, $V_\text{F}$, focusing on a field of view four rows away from the lateral boundaries and the entrance of the medium to minimize boundary effects. Moreover, to minimize exit effects, we measure $V_\text{F}$ when the non-wetting fluid first reaches the halfway point of the medium.

 \section{Influence of a gradient on drainage behavior}

\noindent We first test a disordered, gradient-free porous medium. To incorporate a controlled amount of disorder in the medium, we fabricate porous media with pillars whose diameters are chosen from a normal distribution. We choose the pillar diameters such that the pore diameters are also given by a normal distribution with a mean of \SI{111}{\micro\meter} and with a standard deviation  $\approx\sigma$ = \SI{3.25}{\micro\meter}, such that the difference between the maximum and minimum pore diameters equals $2\sigma$. Hence, the quantity $\sigma$ provides a measure of the disorder in the medium. Because $\text{Ca}\ll1$, we expect the drainage to proceed via capillary fingering. Consistent with this expectation, the fluid displacement proceeds one pore invasion at a time, resulting in a ramified and disordered pathway with $V_\text{F}\approx$ 0.45 as shown in the rightmost panel of Fig. 1. This pathway is morphologically similar to previous observations of capillary fingering pathways, which can have $V_\text{F}$ ranging from $\approx 0.3$ to 0.7 \cite{lenormand1988numerical,chen2017}.

We next test the influence of a pore size gradient on the fluid displacement. To define a gradient, we fabricate porous media with pillars whose diameters, averaged across each row, increase by a fixed amount $|g|$ per row along the imposed flow direction. As a result, the mean pore diameter decreases by $|g|$ from one row to the next, as schematized in the top row of Fig.~\ref{fig:disorder} \cite{chung2017enhancing, chaouche1994invasion,rabbani2018suppressing}. The mean pore diameter in row $n$ is thus given by $\bar{a}_{n}=\bar{a}_{1}+g(n-1)$, where the parameter $g$ quantifies the gradient; in general, a negative or positive value of $g$ indicates that the pore diameters are decreasing or increasing along the flow direction, respectively. We maintain a controlled amount of disorder in the medium by choosing the size of each pore in a given row $n$ from a normal distribution of sizes centered around $\bar{a}_{n}$ and with the difference between the maximum and minimum pore diameters again equal to $2\sigma$. To isolate the influence of the gradient, we fix $\sigma$ = \SI{3.25}{\micro\meter} and test four different values of $g$: $\pm$\SI{10}{\micro\meter} and $\pm$\SI{16.7}{\micro\meter}.

\begin{figure}[b!]
\includegraphics[width=\textwidth]{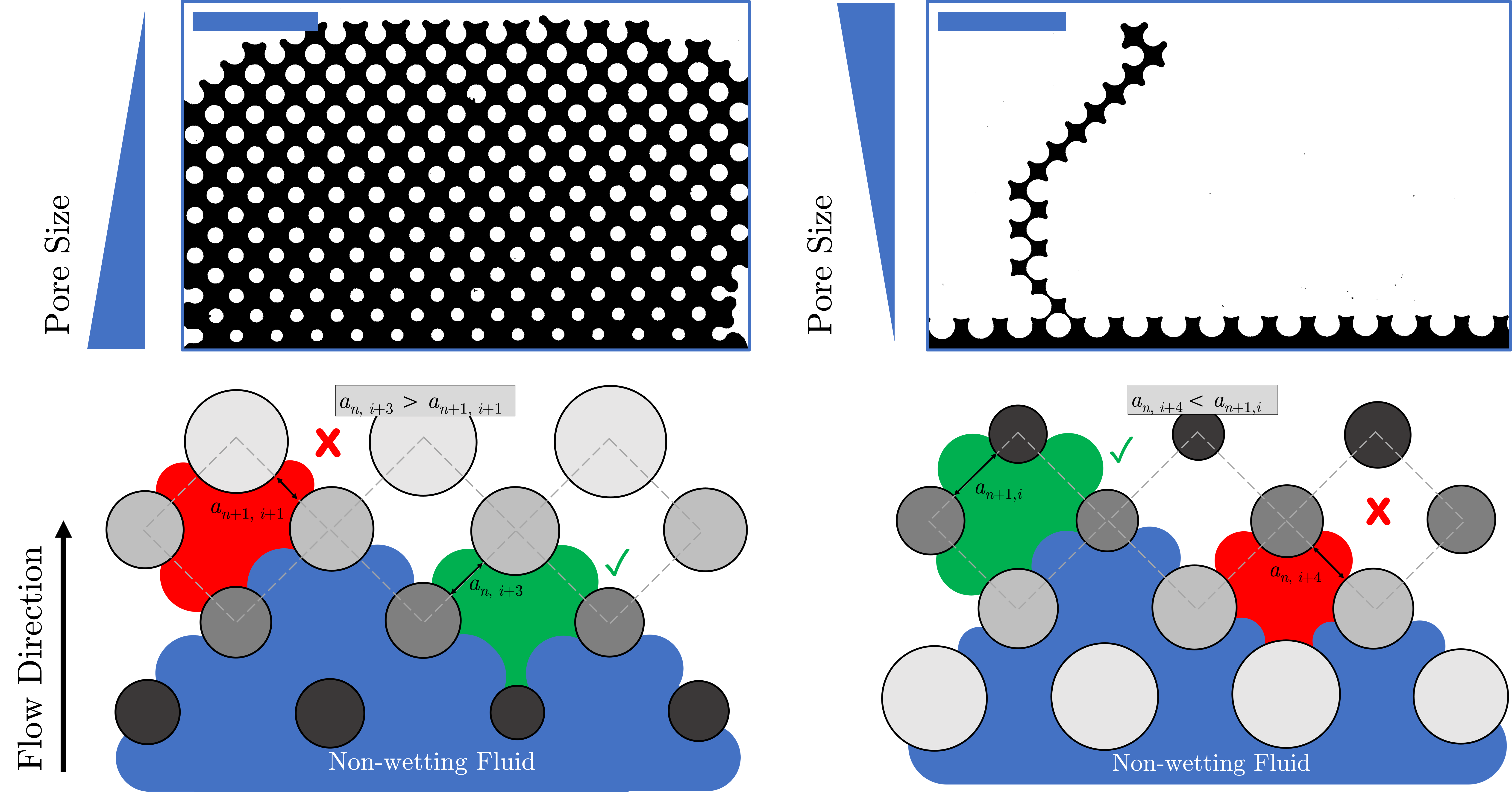}
  \caption{Influence of a gradient on drainage behavior. (Left) Top is an experimental image showing the non-wetting fluid (black) propagating via a uniform, stabilized front in a medium with a gradient of decreasing pore sizes along the flow direction. The capillary pressure threshold is lower for pores in row $n$ compared to pores in row $n+1$; thus, the non-wetting fluid fills all pores in the former row (green in lower panel) before proceeding to the next (red in lower panel). (Right) Top is an experimental image showing the non-wetting fluid (black) propagating via a single, unstable, fingered channel in a medium with a gradient of increasing pore sizes along the flow direction. The capillary pressure threshold is higher for pores in row $n$ compared to pores in row $n+1$; thus, the non-wetting fluid preferentially fills a pore in the next row (green in lower panel) before filling adjacent pores in the current row (red in lower panel). Scale bars are 5 mm. Experimental images show the displacement pathway when the non-wetting fluid first reaches the halfway point of the medium. }
 \label{fig:physics}
\end{figure}

Remarkably, although $\text{Ca}\ll1$, well within the capillary fingering regime for homogeneous media, the addition of the pore size gradient completely suppresses capillary fingering. Instead, we observe two distinct flow pathways. In both cases, the non-wetting fluid displaces the wetting fluid sequentially, one pore invasion at a time; however, the macroscopic flow pathway is starkly different from the ramified and disordered pathway characteristic of capillary fingering. For $g<0$, the non-wetting fluid propagates via a uniform, stabilized front, as shown in the left panel of Fig.~\ref{fig:physics}, ultimately yielding a non-wetting fluid volume fraction $V_\text{F}\approx0.9$. We term this displacement behavior \textit{stable displacement}. By contrast, for $g>0$, the non-wetting fluid propagates through a single, unstable, fingered channel, as shown in the right panel of Fig.~\ref{fig:physics}, ultimately yielding a non-wetting fluid volume fraction $V_\text{F}\approx0.1$. We term this displacement behavior \textit{unstable fingering}. Interestingly, these displacement behaviors arise for both values of $|g|$, and the final $V_\text{F}$ appears to only depend on the sign of $g$, not its magnitude, in this range of $|g|$. 

We can understand the origin of these displacement behaviors by analyzing the pore-scale capillary forces for a fluid interface trapped within a pore in a given row. Because $\text{Ca}\ll1$, the non-wetting fluid pressure must exceed the capillary pressure threshold $\sim\gamma/a$ to enter a pore of diameter $a$. For the media with a pore size gradient, the smallest pore in row $n$, characterized by the largest capillary pressure threshold, has diameter $\bar{a}_{n}-\sigma$; by contrast, the largest pore in row $n+1$, with the smallest capillary pressure threshold, has diameter $\bar{a}_{n+1}+\sigma$. Because $\bar{a}_{n+1}-\bar{a}_{n}=-|g|$, the smallest pore in row $n$ is larger than the largest pore in row $n+1$ when $|g|/2\sigma>1$, as in the experiments, which have $|g|/2\sigma\approx1.5$ and 2.6. Hence, when this macroscopic criterion holds---independent of the exact value of $|g|$---the pores along each row remain separated in diameter from those in the adjacent rows. Consequently,  the capillary pressure threshold is lower for pores in row $n$ compared to pores in row $n+1$. The pore size gradient therefore dominates over disorder in determining the non-wetting fluid pathway: the non-wetting fluid fills all pores in a given row before proceeding to the next, ultimately leading to stable displacement as observed experimentally (Fig.~\ref{fig:physics}, left). 
 
A converse argument holds for the case of $g>0$: when $|g|/2\sigma>1$, the \textit{largest} pore in a given row $n$ is smaller than the \textit{smallest} pore in row $n+1$, independent of the exact value of $|g|$. The magnitudes of the corresponding capillary pressure thresholds again remain separated between adjacent rows; the pressure threshold is then \textit{higher} for pores in row $n$ compared to pores in row $n+1$. The pore size gradient thus dominates over disorder in determining the non-wetting fluid pathway again: in this case, instead of laterally filling a given row, the non-wetting fluid successively fills neighboring pores in adjacent rows \textit{along} the flow direction. As a result, the fluid propagates through the medium in a single thin channel approximately one pore wide, ultimately leading to unstable fingering as observed experimentally (Fig.~\ref{fig:physics}, right).

 \section{Competition between a gradient and disorder} 

\begin{figure*}[t!]
\includegraphics[width=\textwidth]{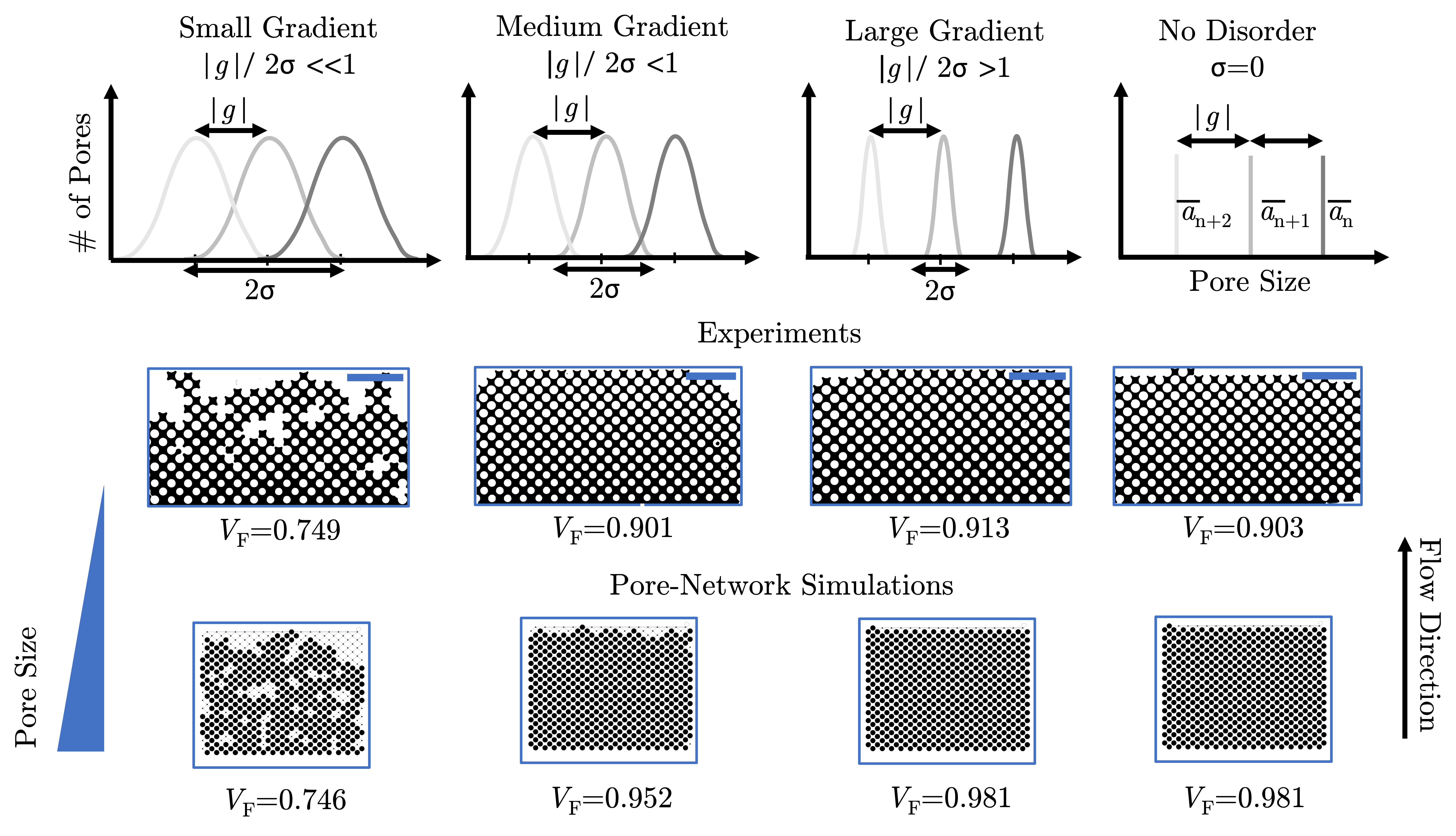}
  \caption{Influence of disorder on drainage behavior. (Top row) Schematic of pore size distributions for each row, indexed by $n$, $n+1$, $n+2$, in porous media with different amounts of disorder. (Middle row) Experimental images and (Bottom row) pore-network simulation images for $g=-\SI{10}{\micro\meter}$ and $2\sigma$ = 130, 20, 2, and \SI{0}{\micro\meter} (left to right). Black shows non-wetting fluid. Scale bars are 5 mm. Experimental images show the displacement pathway when the non-wetting fluid first reaches the halfway point of the medium. For media with a small relative gradient (first column), the pore size distributions for adjacent rows strongly overlap, and the non-wetting fluid pathway is ramified and non-uniform. As the relative gradient increases (second column), the pore size distributions for adjacent rows overlap less, and the non-wetting fluid propagates via a more uniform front. For media with a large relative gradient or no disorder (third and fourth columns), the pore size distributions for adjacent rows do not overlap, and the non-wetting fluid propagates via stable displacement. The corresponding non-wetting fluid volume fraction $V_\text{F}$ increases from the capillary fingering value.}
 \label{fig:disorder}
\end{figure*}

\noindent This geometric argument predicts that the fluid displacement pathway only depends on the sign of $g$, not its magnitude, when the non-dimensional parameter $|g|/2\sigma>1$---resulting in stable displacement or unstable fingering, as observed in the experiments. The threshold $|g|/2\sigma=1$ accordingly represents a macroscopic geometric criterion describing when capillary fingering is completely suppressed, even under conditions in which capillary fingering would typically be predicted for homogeneous disordered media. By contrast, the fluid displacement must proceed by capillary fingering in the limit that $|g|/2\sigma$ approaches zero, in which pore size disorder solely determines the displacement pathway. We thus expect a transition between these displacement behaviors as $|g|/2\sigma$ increases from zero: the pore size distributions and the corresponding capillary pressure thresholds characterizing adjacent rows overlap less and less (top row of Fig.~\ref{fig:disorder}, left to right). The gradient then plays an increasing role in determining the displacement pathway, with stable displacement or unstable fingering increasingly dominating for increasing values of $|g|/2\sigma$, depending on the sign of $g$, up to the threshold $|g|/2\sigma=1$. 

We quantitatively test this prediction by performing experiments on devices with different values of $g$ and $\sigma$ spanning over three decades in the non-dimensional parameter $|g|/2\sigma$. Representative images showing the morphology of the fluid displacement pathway for $g<0$ are shown in the middle row of Fig.~\ref{fig:disorder}. Consistent with our hypothesis, we find that as the relative magnitude of the gradient increases, the non-wetting fluid displacement becomes increasingly uniform for $g<0$, ultimately leading to a stabilized front (Fig.~\ref{fig:disorder}, left to right). Conversely, for $g>0$, we find that as the relative magnitude of the gradient increases, the fluid displacement becomes increasingly unstable, ultimately leading to propagation through a single fingered channel. 

\begin{figure}[htp]
\includegraphics[width=\textwidth]{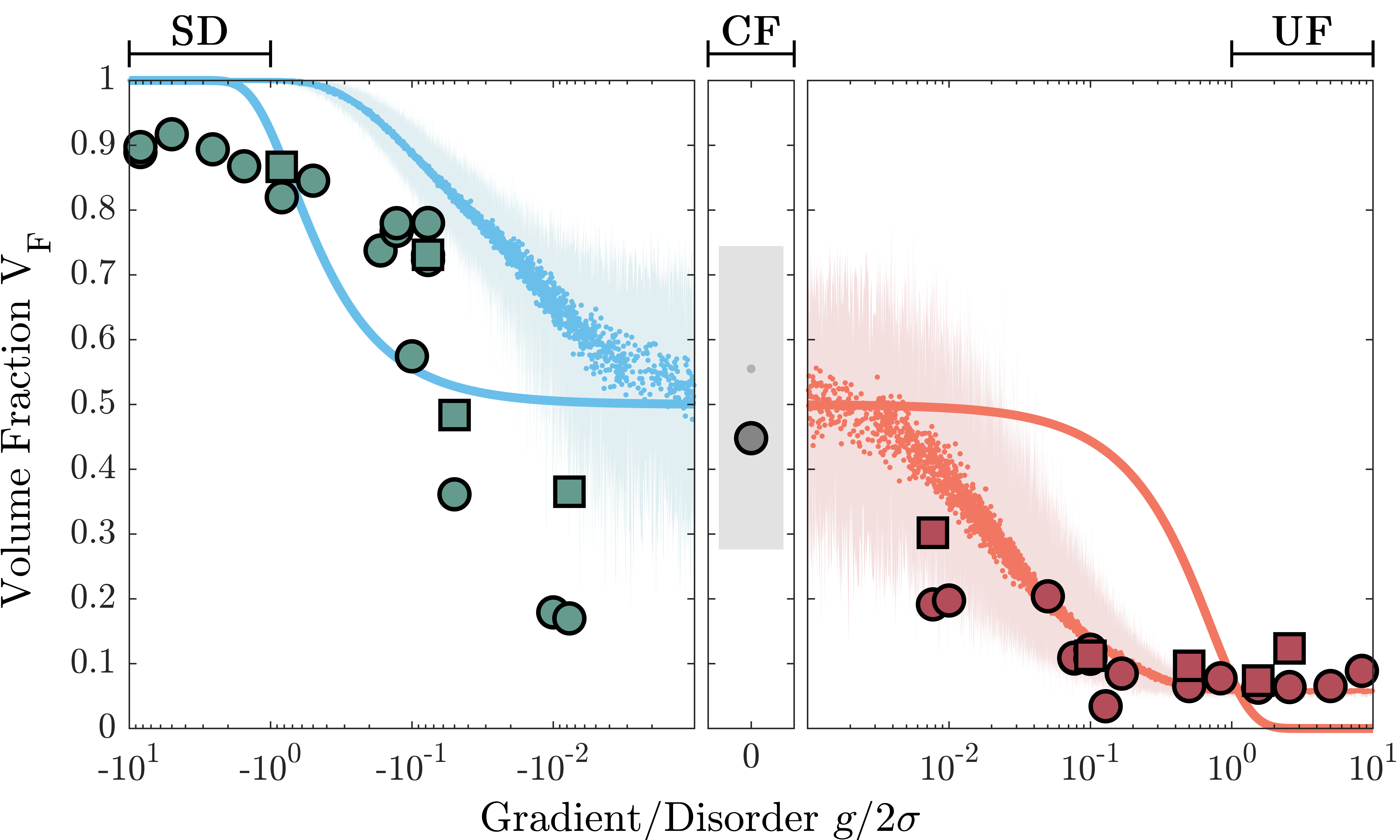}
\caption{Transition from capillary fingering (CF) to stable displacement (SD) or unstable fingering (UF) with an increasing relative gradient.  Large filled symbols show experimental measurements of the non-wetting fluid volume fraction $V_\text{F}$; circles are for measurements taken when the non-wetting fluid first reaches the halfway point of the medium, while square are for measurements taken when the non-wetting fluid breaks through to the outlet. Small filled symbols show results obtained from pore-network model simulations; each point represents the mean value obtained from multiple independent runs having the same $|g|/2\sigma$, which span the full range indicated by the light shaded regions. Solid lines show the solution to Eq. \ref{eq:analytic} derived from a simple geometric argument. The color indicates the sign of $g$; blue represents $g<0$, red represents $g>0$, and grey represents $g=0$. The bars above the plot demarcate the different flow regimes.}
 \label{fig:universalcurve}
\end{figure}

The full set of our measurements of $V_\text{F}$ are shown by the large symbols in Fig.~\ref{fig:universalcurve}. Consistent with our hypothesis, $V_\text{F}$ increases or decreases as $|g|/2\sigma$ increases above zero for $g<0$ and $g>0$, respectively. Surprisingly, these transitions are highly sensitive to the magnitude of the gradient: the non-wetting fluid volume fraction begins to appreciably deviate from the capillary fingering value when $|g|/2\sigma$ is as small as $\sim10^{-2}$, as shown in Fig.~\ref{fig:universalcurve}. Moreover, consistent with our hypothesis, the flow pathway eventually reaches the stable displacement and unstable fingering limits---which we measure to have $V_\text{F}\approx0.9$ and $\approx0.1$---for $g<0$ and $g>0$ respectively, as $|g|/2\sigma$ increases above one. We confirm these limiting values by also testing media with a pore size gradient ($|g|=\SI{10}{\micro\meter}$) but no disorder ($\sigma=0$). Additionally, to confirm that the results are insensitive to the choice of $\text{M}$ in this regime of $\text{Ca}\ll1$, we test different viscosity ratios $\text{M}=4$ and 40 using $65.6$ and $89.1$ vol$\%$ glycerol as the non-wetting fluid, respectively. In both cases, the gradient completely suppresses capillary fingering, as exemplified for the case of $g<0$ in Fig.~\ref{fig:disorder}, and we measure $V_\text{F}=0.932$ and $0.893$ for $g=\SI{-10}{\micro\meter}$ and $V_\text{F}=0.166$ and $0.161$ for $g=\SI{10}{\micro\meter}$, respectively.  Finally, we find qualitatively similar results using different image analysis protocols as indicated by the squares and circles in Fig.~\ref{fig:universalcurve}, further indicating that our findings are robust. The slight discrepancy between the measured volume fraction in the stable displacement limit and the expected value of $V_\text{F}=1$ arises from experimental boundary effects. These are exemplified in Fig. \ref{fig:physics} (Left) and in Fig. \ref{fig:disorder} (Middle Row), which show unfilled pores that occur at the boundary of the medium.

To describe the variation of $V_\text{F}$ with $|g|/2\sigma$, we extend the geometric argument presented in Section II. We again analyze the pore-scale capillary forces for a fluid interface trapped within a pore in row $n$, which represents the downstream tip of the non-wetting fluid interface. However, to more generally represent the pore structure, we describe the distribution of pore diameters $a_{n}$ in the row by the probability density function $p_{n}$. We denote the minimum and maximum pore diameters in row $n$ by $a_{n,\text{min}}$ and $a_{n,\text{max}}$, and the cumulative distribution function of $p_{n}$ by $\Phi_{n}$; $p_{n}(a_{n})$ then represents the probability that a pore in row $n$ has size $a_{n}$, while $\Phi_{n+1}(a_{n})$ represents the fraction of pores in row $n+1$ that are smaller than $a_{n}$. For a gradient-free medium, all the $p_{n}$ are the same and are thus independent of $n$, while as $|g|/2\sigma$ increases from zero, $p_{n}$ and $p_{n+1}$ begin to overlap less and less. We assume that the non-wetting fluid randomly samples pores as it invades the pore space, and that pore invasion only occurs at the downstream tip of the fluid interface. The non-wetting fluid volume fraction can then be approximated by the probability that a pore randomly selected from row $n$ is larger than a pore randomly and independently selected from row $n+1$:
\begin{equation} \label{eq:int}
	V_\text{F}\approx\int_{a_{n,\text{min}}}^{a_{n,\text{max}}}p_{n}(a_{n}) \Phi_{n+1}(a_{n}) \dd a_{n} 
\end{equation}
\noindent For comparison with the experiments, we represent $p_{n}$ by the normal distribution and thereby obtain a full analytic prediction for the non-wetting fluid volume fraction: 
%
%
%
%
%
\begin{equation} \label{eq:analytic}
	V_{F}\approx\frac{1}{\sqrt{2\pi}}\int_{a_{n,\text{min}}}^{a_{n,\text{max}}}\Big[1+\erf\Big(\frac{\xi-g/2\sigma}{\sqrt{2}/2}\Big)\Big]e^{-2\xi^2}\dd\xi 
\end{equation}
\noindent Here, $\xi$ is an integration variable and thus, Equation \ref{eq:analytic} demonstrates the central role of the non-dimensional parameter $g/2\sigma$ in describing the flow dynamics. We numerically solve Eq. \ref{eq:analytic} by replacing $a_{n,\text{min}}$ and $a_{n,\text{max}}$ by $-\infty$ and $+\infty$, respectively. This solution is shown by the solid lines in Fig.~\ref{fig:universalcurve}. 

The comparison between the solution to Eq. \ref{eq:analytic} and the experimental data indicates that our geometric argument qualitatively captures the transition from capillary fingering at low $|g|/2\sigma$ to stable displacement ($g<0$) or unstable fingering ($g>0$) as $|g|/2\sigma$ increases to above one. However, this argument underpredicts the sensitivity to the magnitude of the gradient: the predicted $V_\text{F}$ begins to appreciably deviate from the capillary fingering value only when $|g|/2\sigma\sim10^{-1}$, an order of magnitude larger than in the experiments. This discrepancy reflects the simplifying assumptions made in this model---specifically, that the non-wetting fluid randomly samples pores as it invades the pore space, and that pore invasion only occurs at the downstream tip of the fluid interface. These assumptions neglect the sizes of and the connectivity between  \textit{individual} adjacent pores; they also neglect the possibility of pore invasion upstream from the downstream tip of the fluid interface, as well as wetting fluid trapping throughout the pore space, which decreases the accessibility of pores. Fully describing the influence of a pore size gradient therefore requires a more accurate description of the pore space structure and the pore-scale displacement dynamics.

 \section{Pore-scale computational model} 

\noindent We hypothesize that considering pore-scale capillary forces can describe the transition to capillary fingering with increasing disorder, but only when the sizes and connectivity between individual pairs of pores in the full medium are taken into account. To test this idea, we use a computational pore-network model of invasion percolation with trapping that explicitly considers the pore-to-pore variation in capillary pressure thresholds for a specified medium geometry \cite{lenormand1985invasion, birovljev1991gravity, masson2016fast, wilkinson1986percolation, yortsos2001delineation, xu1998invasion}. 


We first generate a grid of pores arranged on a square diagonal lattice with the same structure as the experimental devices. Using this grid, we use a graph object for the pore network. Pore bodies become nodes that are either ``filled" by the non-wetting fluid or are ``not filled"---that is, filled with the wetting fluid instead---and pore throats become edges connecting two bodies/nodes. The pore body and throat sizes are generated statistically, using specified values of $\sigma$ and $g$, in a manner similar to the experiments. Importantly, the pore throat sizes define the capillary pressure thresholds such that larger pore throats are more easily invaded than smaller ones due to the resultant differences in their capillary pressure thresholds.

At the beginning of each simulation, all the pores in the first row are filled with the non-wetting fluid. At each time step, we determine the connected component clusters of ``not filled" pore bodies; the boundaries with these clusters delineate the invading non-wetting fluid interface or trapped wetting fluid regions. We then find the largest pore throat along the invading non-wetting fluid interface and fill the corresponding pore body, keeping trapped wetting fluid regions unchanged to model an incompressible fluid. The model then iterates through time steps until the non-wetting fluid reaches the outlet. Similar to the experiments, we characterize the resultant flow pathway by calculating the fraction of the pore space that is occupied by the non-wetting fluid, $V_\text{F}$, including both pore bodies and pore throats as in the experiments.

The non-wetting fluid pathways obtained using the pore-network simulations show excellent agreement with the experimental observations. Similar to the experimental observations described in Fig.~\ref{fig:disorder}, we find that for $g<0$ and increasing $|g|/2\sigma$, the non-wetting fluid displacement becomes increasingly uniform for $g<0$, ultimately leading to a stabilized front as shown in the bottom row of Fig.~\ref{fig:disorder}. Moreover, the flow pathway is insensitive to the choice of $\sigma$ for $|g|/2\sigma>1$. Conversely, for $g>0$, we find that as the relative magnitude of the gradient increases, the fluid displacement becomes increasingly unstable, ultimately leading to propagation through a single fingered channel. 

As a final test of our hypothesis, we run pore-network simulations for geometries with different values of $g$ and $\sigma$ spanning over three decades in $|g|/2\sigma$, similar to the experiments. The mean value of $V_\text{F}$ obtained for each $|g|/2\sigma$ is shown by the small symbols in Fig.~\ref{fig:universalcurve}, while the full range of $V_\text{F}$ determined in the simulations is shown by the light shaded regions. As observed in the experiments, as $|g|/2\sigma$ increases, $V_\text{F}$ quickly varies from the two-dimensional capillary fingering value $\approx0.5$, appreciably increasing ($g<0$) or decreasing ($g>0$) when $|g|/2\sigma\sim10^{-2}$. Moreover, as observed in the experiments, the flow pathway eventually reaches the stable displacement and unstable fingering limits $V_\text{F}\approx1.0$ and $\approx0.05$ for $g<0$ and $g>0$, respectively, as $|g|/2\sigma$ increases above 1.  The pore-network simulations thus establish that the macroscopic geometric criterion $|g|/2\sigma=1$ (dashed line in Fig.~\ref{fig:universalcurve}), which does not explicitly consider the pore-scale connectivity of the medium or the full pore invasion dynamics, can still describe when capillary fingering is completely suppressed. 

Intriguingly, the values of $V_\text{F}$ obtained in the simulations are highly variable, similar to the experiments, which show a considerable amount of scatter. We do not observe any apparent correlation between $V_\text{F}$ and the value of $g$ or $\sigma$ used in a simulation (Appendix, Fig. \ref{fig:PCC}), confirming the governing role of the non-dimensional parameter $|g|/2\sigma$ in describing the displacement behavior. Instead, the variability in $V_\text{F}$ reflects the different possible pore structures that can be generated for a given value of $|g|/2\sigma$; these pore structures individually determine the different non-wetting fluid displacement pathways. Given this variability, we find good agreement between the experimental measurements and the pore-network simulations. Moreover, the variability in $V_\text{F}$ decreases as $|g|/2\sigma$ increases for both the experiments and the simulations: the pore size gradient increasingly dominates over disorder in determining the non-wetting fluid pathway. Together, these results support our hypothesis that for $|g|/2\sigma>1$, pore-scale capillary forces can describe the non-wetting fluid pathway, but only when the full pore space structure and the pore-scale displacement dynamics are taken into account.

\section{Conclusion}
\noindent Using experiments with microfluidic porous media and simulations with pore-network models, we demonstrate how capillary fingering is influenced by the competition between a pore size gradient and pore-scale disorder. We find that as the non-dimensional parameter $|g|/2\sigma$ increases, the non-wetting fluid displacement behavior quickly transitions away from the gradient-free limit of capillary fingering, even under flow conditions in which capillary fingering would typically be predicted for homogeneous disordered media. Capillary fingering is eventually completely suppressed when $|g|/2\sigma=1$, above which the non-wetting fluid flows via either stable displacement or unstable fingering, depending on the direction of the gradient. By analyzing capillary forces at the pore scale, we develop a geometric criterion to predict these limiting behaviors. Indeed, because the displacement behaviors studied here are controlled by capillary forces at the pore scale, we expect that our results are also applicable to porous media with wettability gradients. This work thus expands our understanding of drainage in porous media, which is typically described only by the two non-dimensional parameters, $\text{M}$ and $\text{Ca}$ \cite{lenormand1988numerical}. It also suggests a way to control flow behavior in homogeneous porous media through the introduction of a pore size or wettability gradient---for example, via the controlled deposition of solute through the pore space \cite{bradford2008}.

\begin{acknowledgments}
\noindent Correspondence should be addressed to S.S.D. at \texttt{ssdatta@princeton.edu}. It is a pleasure to acknowledge N. Bizmark and H. J. Cho for helpful feedback on the manuscript, and R. K. Prud'homme and H. A. Stone for stimulating discussions. We acknowledge use of the Princeton Institute for the Science and Technology of Materials (PRISM) cleanroom, and the Princeton Institute for Computational Science and Engineering for computer cluster access. This work was supported by start-up funds through Princeton University. N.B.L. was supported in part by the Mary and Randall Hack Graduate Award of the Princeton Environmental Institute. This material is also based upon work supported by the National Science Foundation Graduate Research Fellowship Program (to C.A.B.) under Grant No. DGE-1656466. Any opinions, findings, and conclusions or recommendations expressed in this material are those of the authors and do not necessarily reflect the views of the National Science Foundation.

 \end{acknowledgments}

\begin{appendices}
\counterwithin{figure}{section}
\section{Influence of gradient and disorder on $V_{\text{F}}$}
\noindent To verify that $g/2\sigma$ is an appropriate nondimensional parameter, we show that the volume fraction $V_\text{F}$ obtained in the simulations does not independently depend on $g$ or $\sigma$ for a given value of $g/2\sigma$. This finding is exemplified in Figure \ref{fig:PCC}: $V_\text{F}$ is not correlated with either $g$ (left column) or $\sigma$ (right column) for three different values of $g/2\sigma=1$, $0.1$, and $0.01$ $\pm~2.5\%$ (rows a, b, and c respectively). Furthermore, we calculate the Pearson correlation coefficient $\rho_{A,B}$, which is zero when variables $A$ and $B$ are uncorrelated and $\pm1$ when they are perfectly correlated or anti-correlated. For columns a, b, and c, we find $\rho_{V_\text{F},g}=0.044$, $-0.009$, and $-0.003$, and $\rho_{V_\text{F},\sigma}=0.049$, $-0.007$, and $-0.001$, indicating a lack of correlation.
\setcounter{figure}{0}    
\begin{figure}[htp]
\includegraphics[width=\textwidth]{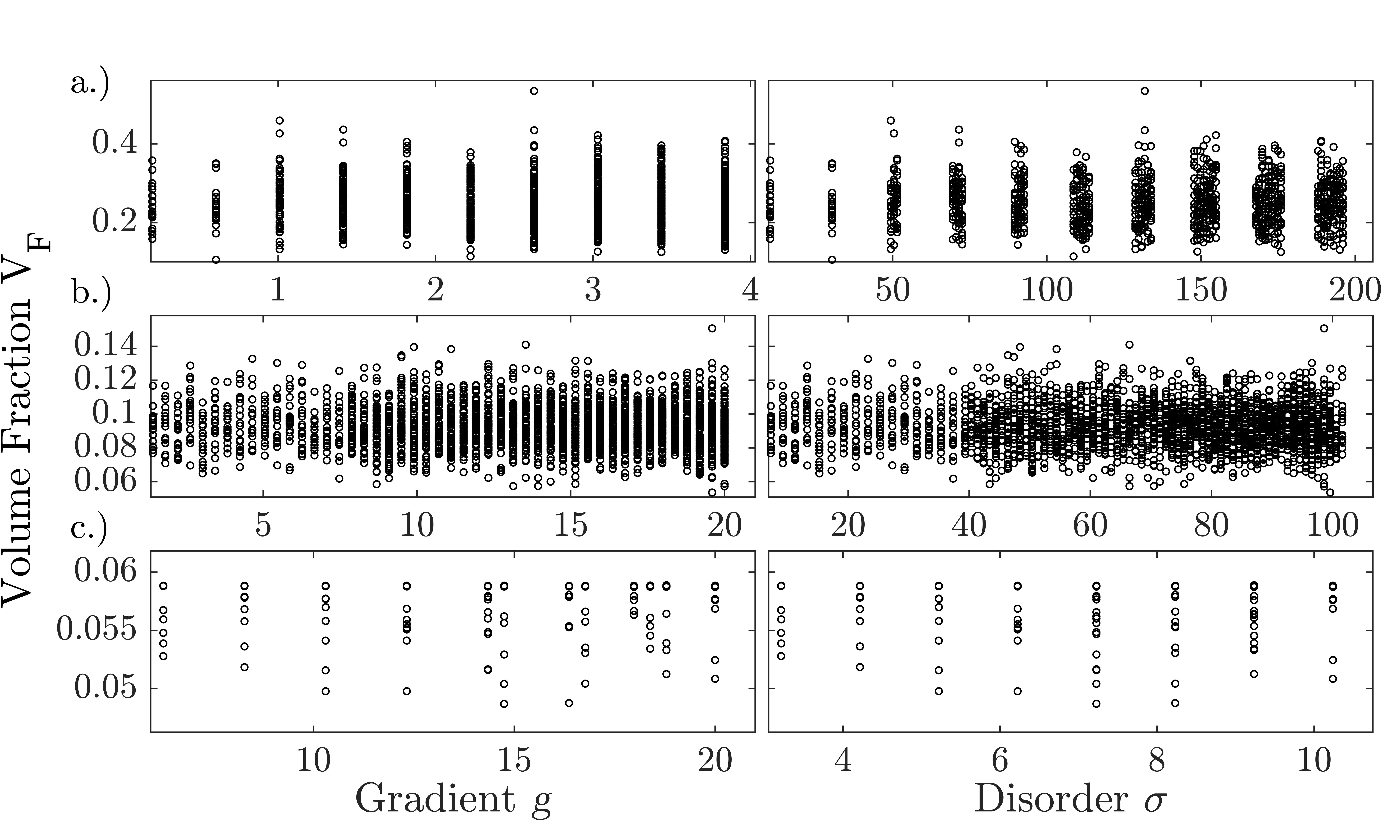}
\caption{Volume fraction $V_F$ as a function of gradient $g$ (left column) or disorder $\sigma$ (right column) in different simulations with $g/2\sigma$ set to (a) 1, (b) 0.1, (c) 0.01 $\pm~2.5\%$. We find no correlation in either case. }
 \label{fig:PCC}
\end{figure}
\end{appendices}
\cleardoublepage

\bibliography{main}

\begin{thebibliography}{10}

\bibitem{cueto2008nonlocal}
L.~Cueto-Felgueroso and R.~Juanes, ``Nonlocal interface dynamics and pattern
  formation in gravity-driven unsaturated flow through porous media,'' {\em
  Physical Review Letters}, vol.~101, no.~24, p.~244504, 2008.

\bibitem{bazyar2018liquid}
H.~Bazyar, P.~Lv, J.~A. Wood, S.~Porada, D.~Lohse, and R.~G. Lammertink,
  ``Liquid--liquid displacement in slippery liquid-infused membranes
  ({S}{L}{I}{M}s),'' {\em Soft Matter}, vol.~14, no.~10, pp.~1780--1788, 2018.

\bibitem{sahimi2011flow}
M.~Sahimi, {\em Flow and {Transport} in {Porous} {Media} and {Fractured}
  {Rock}: {From} classical methods to modern approaches}.
\newblock John Wiley \& Sons, 2011.

\bibitem{berg2012stability}
S.~Berg and H.~Ott, ``Stability of {C}{O}2--brine immiscible displacement,''
  {\em International Journal of Greenhouse Gas Control}, vol.~11, pp.~188--203,
  2012.

\bibitem{bethke1991long}
C.~M. Bethke, J.~D. Reed, and D.~F. Oltz, ``Long-range petroleum migration in
  the {Illinois} {Basin} (1),'' {\em AAPG Bulletin}, vol.~75, no.~5,
  pp.~925--945, 1991.

\bibitem{dandekar2013petroleum}
A.~Y. Dandekar, {\em Petroleum {Reservoir} {Rock} and {Fluid} {Properties}}.
\newblock CRC press, 2013.

\bibitem{benson2008carbon}
S.~M. Benson and F.~M. Orr, ``Carbon dioxide capture and storage,'' {\em MRS
  Bulletin}, vol.~33, no.~4, pp.~303--305, 2008.

\bibitem{macminn2010co}
C.~W. MacMinn, M.~L. Szulczewski, and R.~Juanes, ``C{O}2 migration in saline
  aquifers. {Part} 1. {Capillary} trapping under slope and groundwater flow,''
  {\em Journal of Fluid Mechanics}, vol.~662, pp.~329--351, 2010.

\bibitem{saadatpoor2010new}
E.~Saadatpoor, S.~L. Bryant, and K.~Sepehrnoori, ``New trapping mechanism in
  carbon sequestration,'' {\em Transport in Porous Media}, vol.~82, no.~1,
  pp.~3--17, 2010.

\bibitem{bandara2011pore}
U.~C. Bandara, A.~M. Tartakovsky, and B.~J. Palmer, ``Pore-scale study of
  capillary trapping mechanism during {C}{O}2 injection in geological
  formations,'' {\em International Journal of Greenhouse Gas Control}, vol.~5,
  no.~6, pp.~1566--1577, 2011.

\bibitem{neufeld2009modelling}
J.~A. Neufeld and H.~E. Huppert, ``Modelling carbon dioxide sequestration in
  layered strata,'' {\em Journal of Fluid Mechanics}, vol.~625, pp.~353--370,
  2009.

\bibitem{bear2010modeling}
J.~Bear and A.~H.-D. Cheng, {\em Modeling {Groundwater} {Flow} and
  {Contaminant} {Transport}}, vol.~23.
\newblock Springer Science \& Business Media, 2010.

\bibitem{dawson1997influence}
H.~E. Dawson and P.~V. Roberts, ``Influence of viscous, gravitational, and
  capillary forces on {D}{N}{A}{P}{L} saturation,'' {\em Groundwater}, vol.~35,
  no.~2, pp.~261--269, 1997.

\bibitem{levy2003modelling}
L.~C. Levy, P.~J. Culligan, and J.~T. Germaine, ``Modelling of {D}{N}{A}{P}{L}
  behavior in vertical fractures,'' {\em International Journal of Physical
  Modelling in Geotechnics}, vol.~3, no.~1, pp.~01--18, 2003.

\bibitem{kueper1991two}
B.~H. Kueper and E.~O. Frind, ``Two-phase flow in heterogeneous porous media:
  1. {Model} development,'' {\em Water Resources Research}, vol.~27, no.~6,
  pp.~1049--1057, 1991.

\bibitem{lee2016porous}
C.~H. Lee, R.~Banerjee, F.~Arbabi, J.~Hinebaugh, and A.~Bazylak, ``Porous
  transport layer related mass transport losses in polymer electrolyte membrane
  electrolysis: {A} review,'' in {\em Proceedings of the ASME 2016 14th
  International Conference on Nanochannels, Microchannels, and Minichannels},
  American Society of Mechanical Engineers, 2016.

\bibitem{carmo2013comprehensive}
M.~Carmo, D.~L. Fritz, J.~Mergel, and D.~Stolten, ``A comprehensive review on
  {P}{E}{M} water electrolysis,'' {\em International Journal of Hydrogen
  Energy}, vol.~38, no.~12, pp.~4901--4934, 2013.

\bibitem{rabbani2018suppressing}
H.~S. Rabbani, D.~Or, Y.~Liu, C.-Y. Lai, N.~B. Lu, S.~S. Datta, H.~A. Stone,
  and N.~Shokri, ``Suppressing viscous fingering in structured porous media,''
  {\em Proceedings of the National Academy of Sciences}, vol.~115, no.~19,
  pp.~4833--4838, 2018.

\bibitem{cottin2011influence}
C.~Cottin, H.~Bodiguel, and A.~Colin, ``Influence of wetting conditions on
  drainage in porous media: A microfluidic study,'' {\em Physical Review E},
  vol.~84, no.~2, p.~026311, 2011.

\bibitem{singh2003dynamic}
M.~Singh and K.~K. Mohanty, ``Dynamic modeling of drainage through
  three-dimensional porous materials,'' {\em Chemical Engineering Science},
  vol.~58, no.~1, pp.~1--18, 2003.

\bibitem{datta2014mobilization}
S.~S. Datta, T.~Ramakrishnan, and D.~A. Weitz, ``Mobilization of a trapped
  non-wetting fluid from a three-dimensional porous medium,'' {\em Physics of
  Fluids}, vol.~26, no.~2, p.~022002, 2014.

\bibitem{datta2013drainage}
S.~S. Datta and D.~A. Weitz, ``Drainage in a model stratified porous medium,''
  {\em EPL (Europhysics Letters)}, vol.~101, no.~1, p.~14002, 2013.

\bibitem{lenormand1988numerical}
R.~Lenormand, E.~Touboul, and C.~Zarcone, ``Numerical models and experiments on
  immiscible displacements in porous media,'' {\em Journal of Fluid Mechanics},
  vol.~189, pp.~165--187, 1988.

\bibitem{yortsos1997phase}
Y.~C. Yortsos, B.~Xu, and D.~Salin, ``Phase diagram of fully developed drainage
  in porous media,'' {\em Physical Review Letters}, vol.~79, no.~23, p.~4581,
  1997.

\bibitem{xu1998invasion}
B.~Xu, Y.~Yortsos, and D.~Salin, ``Invasion percolation with viscous forces,''
  {\em Physical Review E}, vol.~57, no.~1, p.~739, 1998.

\bibitem{lenormand1989capillary}
R.~Lenormand and C.~Zarcone, ``Capillary fingering: {Percolation} and fractal
  dimension,'' {\em Transport in Porous Media}, vol.~4, no.~6, pp.~599--612,
  1989.

\bibitem{lenormand1983mechanisms}
R.~Lenormand, C.~Zarcone, and A.~Sarr, ``Mechanisms of the displacement of one
  fluid by another in a network of capillary ducts,'' {\em Journal of Fluid
  Mechanics}, vol.~135, pp.~337--353, 1983.

\bibitem{lenormand1985invasion}
R.~Lenormand and C.~Zarcone, ``Invasion percolation in an etched network:
  {Measurement} of a fractal dimension,'' {\em Physical Review Letters},
  vol.~54, no.~20, p.~2226, 1985.

\bibitem{mayer1965mercury}
R.~P. Mayer and R.~A. Stowe, ``Mercury porosimetry-breakthrough pressure for
  penetration between packed spheres,'' {\em Journal of Colloid Science},
  vol.~20, no.~8, pp.~893--911, 1965.

\bibitem{xu2008dynamics}
L.~Xu, S.~Davies, A.~B. Schofield, and D.~A. Weitz, ``Dynamics of drying in
  3{D} porous media,'' {\em Physical Review Letters}, vol.~101, no.~9,
  p.~094502, 2008.

\bibitem{krummel2013visualizing}
A.~T. Krummel, S.~S. Datta, S.~M{\"u}nster, and D.~A. Weitz, ``Visualizing
  multiphase flow and trapped fluid configurations in a model three-dimensional
  porous medium,'' {\em AIChE Journal}, vol.~59, no.~3, pp.~1022--1029, 2013.

\bibitem{toledo1994pore}
P.~G. Toledo, L.~Scriven, H.~T. Davis, {\em et~al.}, ``Pore-space statistics
  and capillary pressure curves from volume-controlled porosimetry,'' {\em SPE
  Formation Evaluation}, vol.~9, no.~01, pp.~46--54, 1994.

\bibitem{mason1986meniscus}
G.~Mason and N.~Morrow, ``Meniscus displacement curvatures of a perfectly
  wetting liquid in capillary pore throats formed by spheres,'' {\em Journal of
  Colloid and Interface Science}, vol.~109, no.~1, pp.~46--56, 1986.

\bibitem{joekar2012analysis}
V.~Joekar-Niasar and S.~Hassanizadeh, ``Analysis of fundamentals of two-phase
  flow in porous media using dynamic pore-network models: A review,'' {\em
  Critical Reviews in Environmental Science and Technology}, vol.~42, no.~18,
  pp.~1895--1976, 2012.

\bibitem{maaloy1992dynamics}
K.~J. M{\aa}l{\o}y, L.~Furuberg, J.~Feder, and T.~J{\o}ssang, ``Dynamics of
  slow drainage in porous media,'' {\em Physical Review Letters}, vol.~68,
  no.~14, p.~2161, 1992.

\bibitem{martys1991critical}
N.~Martys, M.~Cieplak, and M.~O. Robbins, ``Critical phenomena in fluid
  invasion of porous media,'' {\em Physical Review Letters}, vol.~66, no.~8,
  p.~1058, 1991.

\bibitem{ringrose1993immiscible}
P.~Ringrose, K.~Sorbie, P.~Corbett, and J.~Jensen, ``Immiscible flow behaviour
  in laminated and cross-bedded sandstones,'' {\em Journal of Petroleum Science
  and Engineering}, vol.~9, no.~2, pp.~103--124, 1993.

\bibitem{schaetzl2015soils}
R.~J. Schaetzl and M.~L. Thompson, {\em Soils}.
\newblock Cambridge university press, 2015.

\bibitem{ashraf2019capillary}
S.~Ashraf and J.~Phirani, ``Capillary displacement of viscous liquids in a
  multi-layered porous medium,'' {\em Soft Matter}, vol.~15, no.~9,
  pp.~2057--2070, 2019.

\bibitem{yokoyama1981effects}
Y.~Yokoyama, L.~W. Lake, {\em et~al.}, ``The effects of capillary pressure on
  immiscible displacements in stratified porous media,'' in {\em SPE Annual
  Technical Conference and Exhibition}, Society of Petroleum Engineers, 1981.

\bibitem{lake1981taylor}
L.~W. Lake, G.~J. Hirasaki, {\em et~al.}, ``Taylor's dispersion in stratified
  porous media,'' {\em Society of Petroleum Engineers Journal}, vol.~21,
  no.~04, pp.~459--468, 1981.

\bibitem{chatzis1995investigation}
L.~Chatzis and S.~Ayalollahi, ``Investigation of the {G}{A}{I}{G}{I} process in
  stratified porous media for the recovery of waterflood residual oil,'' in
  {\em Technical Meeting/Petroleum Conference of The South Saskatchewan
  Section}, Petroleum Society of Canada, 1995.

\bibitem{meakin1991}
P.~Meakin, A.~Birovljev, V.~Frette, J.~Feder, and T.~Jossang, ``Gradient
  stabilized and destabilized invasion percolation,'' {\em Physica A},
  vol.~191, pp.~227--239, 1991.

\bibitem{meakin1992}
P.~Meakin, J.~Feder, V.~Frette, and T.~Jossang, ``Invasion percolation in a
  destabilizing gradient,'' {\em Physical Review A}, vol.~46, no.~6,
  pp.~3357--3368, 1992.

\bibitem{chaouche1994}
M.~Chaouche, N.~Rakotomalala, D.~Salin, B.~Xu, and Y.~Yortsos, ``Invasion
  percolation in a hydrostatic or permeability gradient: {E}xperiments and
  simulations,'' {\em Physical Review E}, vol.~49, no.~5, pp.~4133--4139, 1994.

\bibitem{xu1998}
B.~Xu, Y.~Yortsos, and D.~Salin, ``Invasion percolation with viscous forces,''
  {\em Physical Review E}, vol.~57, no.~1, pp.~739--751, 1998.

\bibitem{yortsos2001}
Y.~Yortsos, B.~Xu, and D.~Salin, ``Delineation of microscale regimes of
  fully-developed drainage and implications for continuum models,'' {\em
  Computational Geosciences}, vol.~5, pp.~257--278, 2001.

\bibitem{al2012control}
T.~T. Al-Housseiny, P.~A. Tsai, and H.~A. Stone, ``Control of interfacial
  instabilities using flow geometry,'' {\em Nature Physics}, vol.~8, no.~10,
  p.~747, 2012.

\bibitem{jackson2017stability}
S.~Jackson, H.~Power, D.~Giddings, and D.~Stevens, ``The stability of
  immiscible viscous fingering in {Hele}-{Shaw} cells with spatially varying
  permeability,'' {\em Computer Methods in Applied Mechanics and Engineering},
  vol.~320, pp.~606--632, 2017.

\bibitem{pihler2012suppression}
D.~Pihler-Puzovi{\'c}, P.~Illien, M.~Heil, and A.~Juel, ``Suppression of
  complex fingerlike patterns at the interface between air and a viscous fluid
  by elastic membranes,'' {\em Physical Review Letters}, vol.~108, no.~7,
  p.~074502, 2012.

\bibitem{biswas2018drying}
S.~Biswas, P.~Fantinel, O.~Borgman, R.~Holtzman, and L.~Goehring, ``Drying and
  percolation in spatially correlated porous media,'' {\em Physical Review
  Fluids}, vol.~3, p.~124307, 2018.

\bibitem{chen2017}
Y.~Chen, S.~Fang, D.-S. Wu, and R.~Hu, ``Visualizing and quantifying the
  crossover from capillary fingering to viscous fingering in a rough
  fracture,'' {\em Water Resources Research}, vol.~53, no.~9, pp.~7756--7772,
  2017.

\bibitem{chung2017enhancing}
C.~Chung and H.-Y. Lin, ``Enhancing immiscible fluid displacement in porous
  media by capillary pressure discontinuities,'' {\em Transport in Porous
  Media}, vol.~120, no.~2, pp.~309--325, 2017.

\bibitem{chaouche1994invasion}
M.~Chaouche, N.~Rakotomalala, D.~Salin, B.~Xu, and Y.~Yortsos, ``Invasion
  percolation in a hydrostatic or permeability gradient: Experiments and
  simulations,'' {\em Physical Review E}, vol.~49, no.~5, p.~4133, 1994.

\bibitem{birovljev1991gravity}
A.~Birovljev, L.~Furuberg, J.~Feder, T.~Jssang, K.~Mly, and A.~Aharony,
  ``Gravity invasion percolation in two dimensions: {Experiment} and
  simulation,'' {\em Physical Review Letters}, vol.~67, no.~5, p.~584, 1991.

\bibitem{masson2016fast}
Y.~Masson, ``A fast two-step algorithm for invasion percolation with
  trapping,'' {\em Computers \& Geosciences}, vol.~90, pp.~41--48, 2016.

\bibitem{wilkinson1986percolation}
D.~Wilkinson, ``Percolation effects in immiscible displacement,'' {\em Physical
  Review A}, vol.~34, no.~2, p.~1380, 1986.

\bibitem{yortsos2001delineation}
Y.~C. Yortsos, B.~Xu, and D.~Salin, ``Delineation of microscale regimes of
  fully-developed drainage and implications for continuum models,'' {\em
  Computational Geosciences}, vol.~5, no.~3, pp.~257--278, 2001.

\bibitem{bradford2008}
S.~Bradford and S.~Torkzaban, ``Colloid transport and retention in unsaturated
  porous media: {A} review of interface-, collector-, and pore-scale processes
  and models,'' {\em Vadose Zone Journal}, vol.~7, no.~2, pp.~667--681, 2008.

\end{thebibliography}

\end{document}